\documentclass[review]{elsarticle}
\graphicspath{ {./figures/} }
\usepackage{hyperref}
\usepackage{float}
\usepackage{verbatim} 
\usepackage{apalike}
\restylefloat{figure}
\usepackage{rotating} 

\graphicspath{ {./figures/} }
\usepackage[table]{xcolor}
\usepackage{verbatim} 
\usepackage{apalike}
\usepackage{caption}
\usepackage{subcaption}
\usepackage[margin=1in]{geometry} 
\usepackage[]{graphicx}
\usepackage[font=small,skip=0pt]{caption}
\usepackage{enumitem}
\usepackage{booktabs}
\usepackage{longtable}
\usepackage{tabularx}
\usepackage{cancel}
\usepackage{multirow}
\usepackage{supertabular}
\usepackage{algorithmic}
\usepackage{algorithm}
\usepackage{amsmath}
\usepackage{rotating}
\usepackage[normalem]{ulem}
\usepackage[utf8]{inputenc}
\usepackage[T1]{fontenc}
\usepackage{lineno}
\usepackage{array}
\usepackage{makecell}
\usepackage{pdflscape}
\usepackage{afterpage}
\usepackage{capt-of}
\usepackage{lipsum}
\usepackage{changepage}
\usepackage{xcolor}
\usepackage{rotating} 
\usepackage{graphicx}
\newcommand{\blue}[1]{\textcolor{blue}{#1}}
\usepackage{amssymb}
\usepackage{natbib}
\bibpunct{(}{)}{,}{a}{}{;}
\renewcommand{\citep}[1]{(\citealp{#1})}

\biboptions{authoryear}

\begin{document}

\begin{frontmatter}

\title{ Automated Disease Diagnosis in Pumpkin Plants Using Advanced CNN Models}

\author[label1]{Aymane Khaldi}
\ead{ay.khaldi@aui.ma}
\author[label1]{El Mostafa Kalmoun \corref{cor1}}
\ead{e.kalmoun@aui.ma}

\cortext[cor1]{Corresponding author.}
\address[label1]{School of Science and Engineering, Al Akhawayn University, Ifrane, Morocco}

\begin{abstract}
Pumpkin is a vital crop cultivated globally, and its productivity is crucial for food security, especially in developing regions. Accurate and timely detection of pumpkin leaf diseases is essential to mitigate significant losses in yield and quality. Traditional methods of disease identification rely heavily on subjective judgment by farmers or experts, which can lead to inefficiencies and missed opportunities for intervention. Recent advancements in machine learning and deep learning offer promising solutions for automating and improving the accuracy of plant disease detection.

This paper presents a comprehensive analysis of state-of-the-art Convolutional Neural Network (CNN) models for classifying diseases in pumpkin plant leaves. Using a publicly available dataset of 2000 high-resolution images, we evaluate the performance of several CNN architectures, including ResNet, DenseNet, and EfficientNet, in recognizing five classes: healthy leaves and four common diseases—downy mildew, powdery mildew, mosaic disease, and bacterial leaf spot. We fine-tuned these pretrained models and conducted hyperparameter optimization experiments. ResNet-34, DenseNet-121, and EfficientNet-B7 were identified as top-performing models, each excelling in different classes of leaf diseases.

Our analysis revealed DenseNet-121 as the optimal model when considering both accuracy and computational complexity achieving an overall accuracy of 86\%. This study underscores the potential of CNNs in automating disease diagnosis for pumpkin plants, offering valuable insights that can contribute to enhancing agricultural productivity and minimizing economic losses.
\end{abstract}


\end{frontmatter}

\section{Introduction}\label{S1}
Agriculture is a key component of the global economy, with significant advancements in recent years to meet the rising demand for food driven by population growth. This sector plays a vital role in boosting the economic growth of many countries \citep{bezabh2024classification, mondal2023plant}. 

Plant diseases caused by bacteria, fungi, viruses, mites, and pests significantly reduce crop yields each year, impacting production, growth, and economic stability. Around 40\% of global crop yield is lost annually due to pests, resulting in a financial loss of \$70 billion, while total losses from all plant diseases reach \$220 billion annually \citep{un2021}. Early and accurate detection of these diseases is crucial to prevent crop waste and reduce financial losses \citep{mondal2023plant}.

The identification of crop diseases largely relies on farmers' personal experience, which often lacks objectivity and leads to suboptimal outcomes \citep{han2012measurement}. Expert manual detection of plant diseases is time-consuming, costly, labor-intensive, and may involve destructive methods. Additionally, subjective judgments can result in missed opportunities for timely prevention or treatment \citep{shougang2016segmentation, sankaran2010review}.
To address these limitations, along with recent advancements in computer vision, various machine learning and deep learning models have been applied to plant disease recognition. These approaches enable automated, reliable, and non-destructive detection of plant diseases, helping to enhance and predict agricultural productivity \citep{jackulin2022comprehensive, harakannanavar2022plant, kotwal2023agricultural, dang2023vpbr}.

Pumpkin is a widely cultivated and consumed vegetable across Asia, Europe, and North America \citep{hussain2022utilization}. Its production plays a crucial role in the agricultural goals of food sufficiency in developing countries. Identifying pumpkin diseases early is essential to prevent significant losses in both productivity and crop quality \citep{bezabh2024classification}. However, recognizing pumpkin leaf diseases can be challenging by mere visual inspection. A deep learning approach, leveraging leaf color, can be instrumental in accurately detecting pumpkin plant diseases \citep{mithu2022pumpkin}.

Few studies have utilized machine and deep learning models for automating the recognition of pumpkin leaf diseases. For example, \cite{gomathy2023intelligent} applied a feed-forward neural network classifier to detect pumpkin leaf disease in India, while \cite{mithu2022pumpkin} compared various machine learning models with non-state-of-the-art Convolutional Neural Networks (CNNs) for disease recognition. \cite{bezabh2024classification} developed a hybrid method combining LeNet and ResNet to classify pumpkin leaf and fruit rot diseases, and \cite{sethy2019detection} used k-means and PCA for identifying downy mildew in pumpkin leaves. Despite these efforts, a comprehensive study on the effectiveness of state-of-the-art CNN models for pumpkin leaf disease detection is lacking.

This study aims to conduct a comparative analysis of several state-of-the-art CNN models, including ResNet, DenseNet, and EfficientNet, for classifying pumpkin leaves into healthy and four common diseases—downy mildew, powdery mildew, mosaic disease, and bacterial leaf spot. To the best of our knowledge, this is the first study to explore different CNN architectures for pumpkin leaf disease classification. The rest of the paper is organized as follows: Section \ref{S2} describes the experimental setup, Section \ref{S3} presents the experimental results, and Section \ref{S4} concludes the study and sheds light int future work.

\section{Experiment}\label{S2}
In this section, we present the pumpkin leaf disease dataset, as well as the pipeline used to perform the automatic recognition of leaf disease, finally, we present the metrics used for model evaluation.

\subsection{Data Acquisition}
\begin{figure}[H]
    \centering
    \includegraphics[width=15cm]{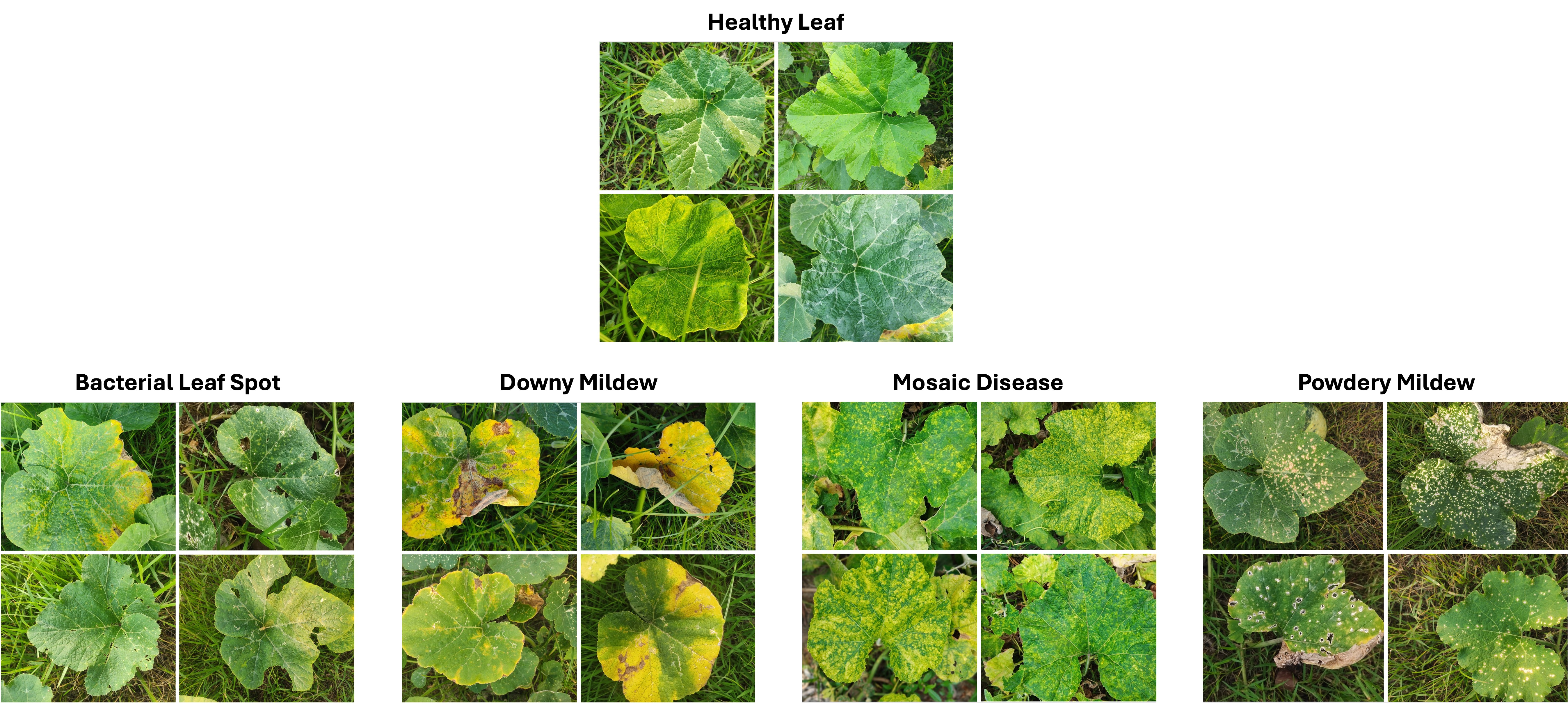}
    \caption {Visualization of four samples per class from the Pumpkin Leaf Diseases dataset.}
    \label{data}
\end{figure}

In this study, we used the publicly available "Pumpkin Leaf Diseases" dataset, which can be accessed from Kaggle via this link \href{https://www.kaggle.com/datasets/tahmidmir/pumpkin-leaf-diseases-dataset-from-bangladesh}{\blue{(Pumpkin Leaf Diseases)}}. The dataset contains 2000 high-resolution RGB images of pumpkin leaves, with 400 images per class. It features both healthy leaves and four categories of diseased leaves: downy mildew, powdery mildew, mosaic disease, and bacterial leaf spot. Representative samples from each class are shown in Fig. \ref{data}. This dataset is intended for the study and identification of various pumpkin leaf diseases and for differentiating between healthy and diseased leaves.

For the experimental analysis, the dataset was divided into three subsets: (1) 80\% as training set to optimize model parameters, (2) 10\% as validation set to determine the best model configuration and prevent overfitting, and (3) 10\% as test set to assess the model's generalization performance.

\subsection{Experimental Design}
In this study, we evaluated different architectures of state-of-the-art CNN models, including ResNet, DenseNet, and EfficientNet, each with varying numbers of layers. All models were pretrained on the ImageNet dataset \citep{deng2009imagenet} and then fine-tuned for the specific task of pumpkin leaf disease classification. To adapt these models for our 5-class problem, the classifier heads were modified accordingly, and the input images were resized to 224x224 and normalized using ImageNet statistics. The overall learning approach is depicted in Fig. \ref{exp}.

\begin{figure}[H]
\centering
\includegraphics[width=12cm]{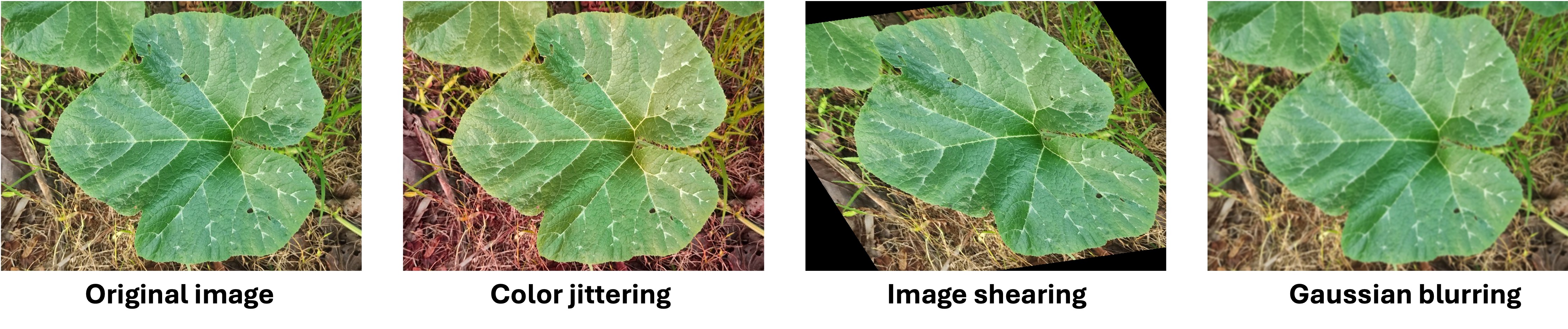}

\caption{Visualization of the three transforms used in data augmentation.}
\label{DA}
\end{figure}

\begin{figure}[H]
\centering
\includegraphics[width=12cm]{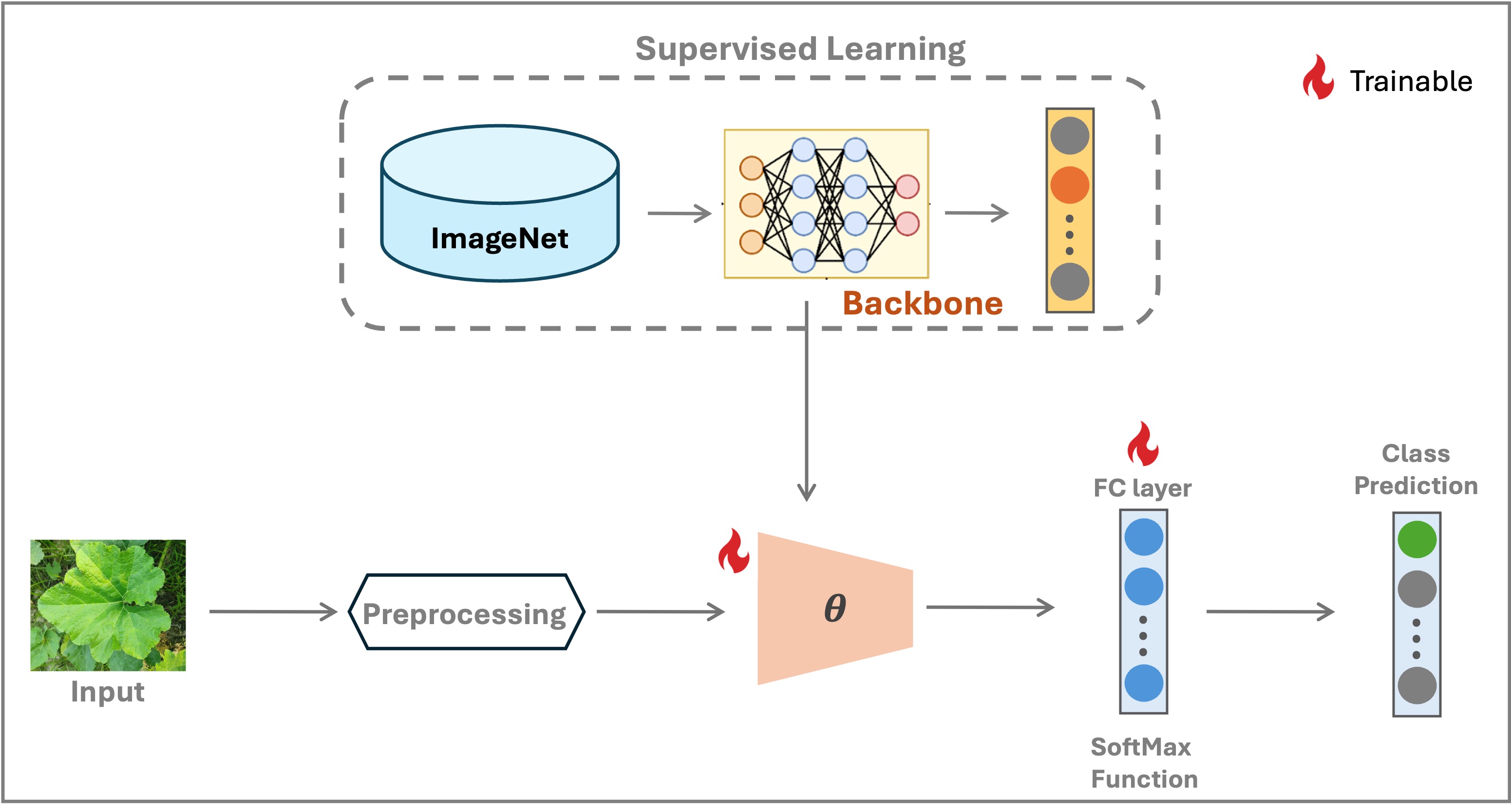}
 
\caption{Workflow describing the learning approach used for pumpkin leaf disease recognition.}
\label{exp}
\end{figure}

To determine the optimal CNN configuration, we initially fixed the backbone to ResNet-50 and explored several hyperparameter settings: (1) Five different learning rates ranging from $10^{-2}$ to $10^{-6}$ with a step size of 10. (2) Two multi-step learning rate schedulers, one decreasing the learning rate by a factor of 10 every 5 epochs and the other every 10 epochs. (3) Five batch sizes: $\{4, 8, 16, 32, 64\}$. (4) The use of data augmentation techniques combining color jittering, image shearing, and Gaussian blurring (Fig \ref{DA}). Each model was trained for 50 epochs using the Adam optimizer with weight decay. To mitigate overfitting, the models were evaluated on the validation set after each epoch, on which the best model was saved. After selecting the optimal configuration, we trained and tested 17 different backbones.

\subsection{Model Evaluation}
To evaluate the overall performance of the model, we employed three metrics: precision (\ref{eq2}) to measure the accuracy of the model in recognizing different pumpkin leaf diseases, recall (\ref{eq3}) to assess the model’s effectiveness in retrieving instances of each disease, and accuracy (\ref{eq1}), which indicates the proportion of correctly classified images relative to the total number of images.

\begin{equation}
    \text{Precision} = \frac{\text{TP}}{\text{TP+FP}}
    \label{eq2}
\end{equation}

\begin{equation}
    \text{Recall} = \frac{\text{TP}}{\text{TP+FN}}
    \label{eq3}
\end{equation}

\begin{equation}
    \text{Accuracy} = \frac{\text{TP+TN}}{\text{TP+FP+TN+FN}}
    \label{eq1}
\end{equation}

Where: TP (True Positives), TN (True Negatives), FP (False Positives), and FN (False Negatives).

\section{Results and Discussion}\label{S3}

\begin{figure}[H]
  \begin{center}
    \includegraphics[width=9cm]{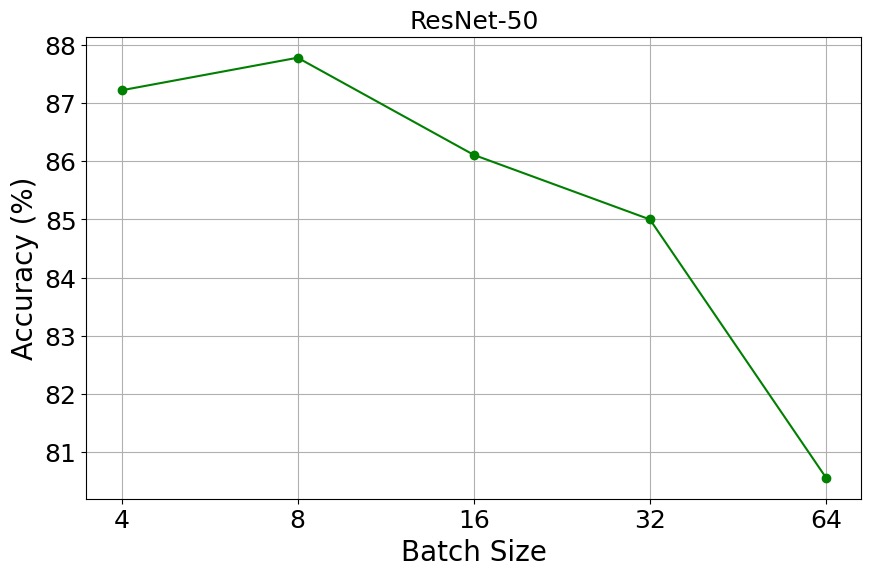} 
  \end{center}
  \caption{Accuracy of the ResNet-50 model evaluated on the validation set across various batch sizes.}
  \label{fig4}
\end{figure}

\begin{figure}[H]
  \begin{center}
    \includegraphics[width=10cm]{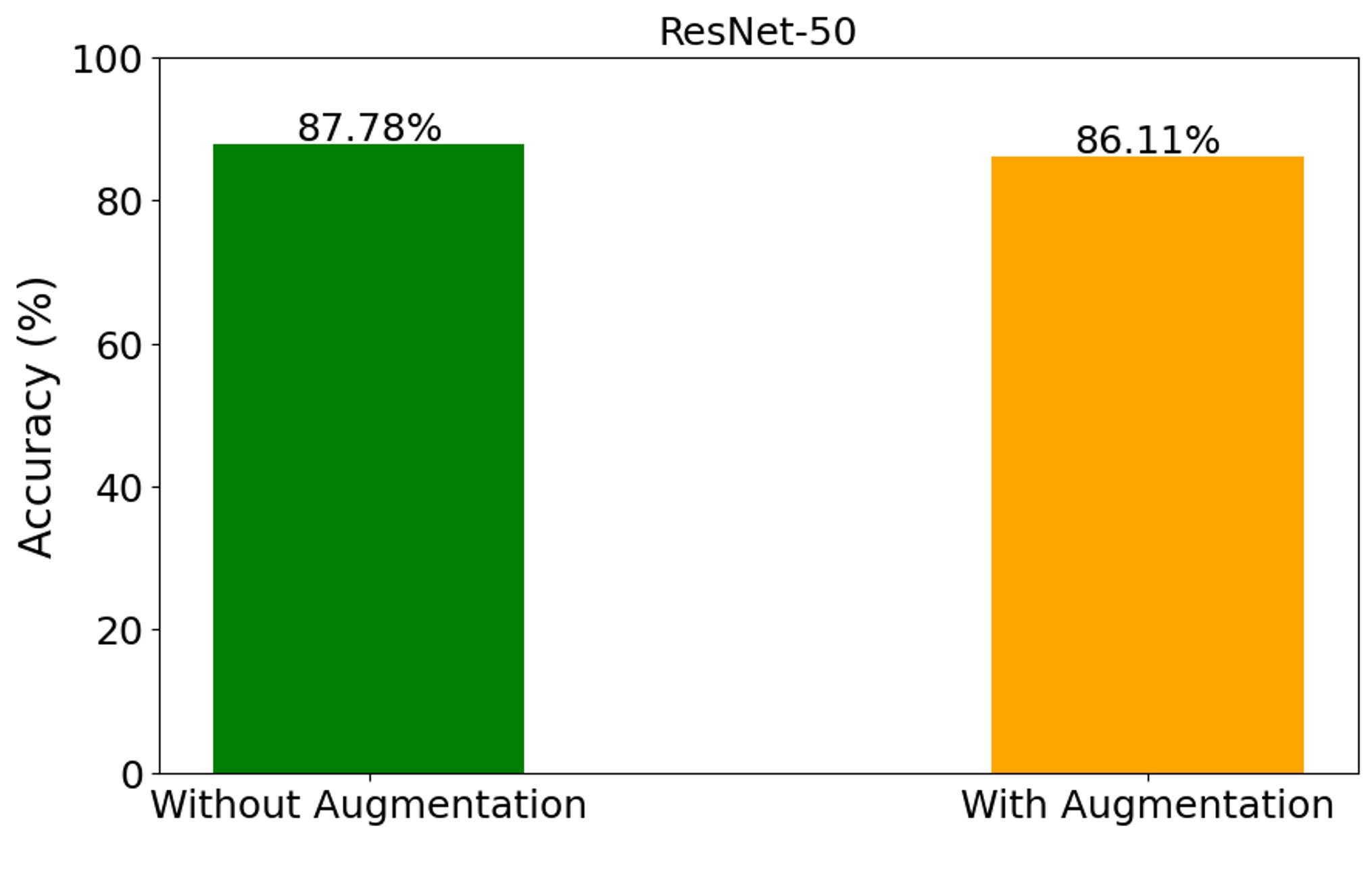} 
  \end{center}
  \caption{Accuracy of the ResNet-50 model evaluated on the validation set and trained with and without data augmentation using a batch size of 8 images.}
  \label{fig5}
\end{figure}

Table \ref{tab1} shows the validation results of ResNet-50 using a multi-step learning rate scheduler with various decay rates and initial learning rates. The best model accuracy, 87.78\%, was achieved with the scheduler decaying every five epochs and with an initial learning rate of $10^{-6}$. Fig. \ref{fig4} illustrates the validation performance of ResNet-50 across different batch sizes (4, 8, 16, 32, 64). The highest accuracy, 87.78\%, was achieved with a batch size of 8 images. Fig. \ref{fig5} compares the validation accuracy of ResNet-50 trained with and without data augmentation. The best performance, 87.78\%, was observed without any augmentation, compared to 86.11\% when augmentation techniques were applied.

\begin{table}[H]
\centering
\caption{Accuracy of the ResNet-50 model evaluated on the validation set across two different learning rate schedulers and various initial values of learning rates using a batch size of 8 images.}
\label{tab1}
\begin{tabular}{lrr}
\hline
\multicolumn{1}{l}{Learning rate scheduler}          & Initial learning rate   & Accuracy (\%) \\ \hline
\multirow{5}{*}{MultiStep after 5 epochs}  & 0.01     & 83.64    \\
                                          & 0.001    & 85.56    \\
                                          & 0.0001   & 86.33    \\
                                          & \textbf{0.00001}  & \textbf{87.78}    \\
                                          & 0.000001 & 53.89    \\ \hline
\multirow{5}{*}{MultiStep after 10 epochs} & 0.01     & 83.89    \\
                                          & 0.001    & 83.89    \\
                                          & 0.0001   & 85.00       \\
                                          & 0.00001  & 86.67    \\
                                          & 0.000001 & 70.56    \\ \hline
\end{tabular}
\end{table}

\begin{figure}[H]
  \begin{center}
    \includegraphics[width=12cm]{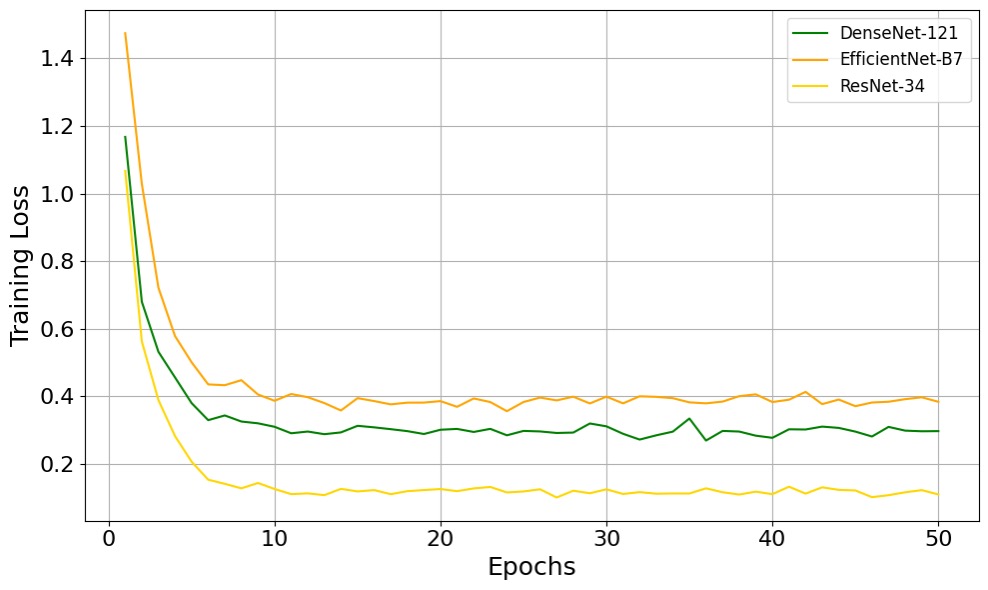} 
  \end{center}
  \caption{Learning curves showing the training process for the top three models: DenseNet-121, EfficientNet-B7, and ResNet-34.}
  \label{fig6}
\end{figure}

Table \ref{tab2} outlines the test results of various CNN backbones—ResNet, DenseNet, and EfficientNet—using the optimal configuration identified in earlier experiments. The top-performing models, ResNet-34, DenseNet-121, and EfficientNet-B7, all achieved an accuracy of 86\%. Despite sharing the same accuracy, their memory complexities differ: EfficientNet-B7 is the most resource-intensive with 5.32 GMac (Giga Multiply-Accumulate Operations) and 66.35 million parameters, followed by ResNet-34 with 3.68 GMac and 21.80 million parameters, and DenseNet-121 with 2.90 GMac and 7.98 million parameters. Fig. \ref{fig6} plots the learning curves of these models over 50 epochs, with ResNet-34 converging faster, followed by DenseNet-121 and then EfficientNet-B7.

\begin{table}[H]
\centering
\caption{Test results for the different evaluated backbones using a batch size of 8 images.}
\label{tab2}
\begin{tabular}{lrrrrr}
\hline
\multicolumn{1}{c}{\textbf{Backbone}} &
  \multicolumn{1}{c}{\textbf{Accuracy}} &
  \multicolumn{1}{c}{\textbf{Precision}} &
  \multicolumn{1}{c}{\textbf{Recall}} &
  \multicolumn{1}{c}{\textbf{GMac}} &
  \textbf{Params (M)} \\ \hline
ResNet-18       & 84.00 & 84.43 & 84.00 & 1.83  & 11.69 \\
\textbf{ResNet-34}       & \textbf{86.00} & \textbf{86.81} & \textbf{86.00} & \textbf{3.68}  & \textbf{21.80} \\
ResNet-50       & 84.50 & 85.05 & 84.50 & 4.13  & 25.56 \\
ResNet-101      & 84.00 & 84.22 & 84.00 & 7.87  & 44.55 \\
ResNet-152      & 85.50 & 86.10 & 85.50 & 11.60 & 60.19 \\
\textbf{DenseNet-121}    & \textbf{86.00} & \textbf{86.33} & \textbf{86.00} & \textbf{2.90}  & \textbf{7.98}  \\
DenseNet-161    & 85.50 & 85.48 & 85.50 & 7.85  & 28.68 \\
DenseNet-169    & 85.00 & 85.29 & 85.00 & 3.44  & 14.15 \\
DenseNet-201    & 85.00 & 85.39 & 85.00 & 4.39  & 20.01 \\
EfficientNet-B0 & 82.50 & 83.39 & 82.50 & 0.41  & 5.29  \\
EfficientNet-B1 & 81.00 & 81.32 & 81.00 & 0.60  & 7.79  \\
EfficientNet-B2 & 84.50 & 85.17 & 84.50 & 0.69  & 9.11  \\
EfficientNet-B3 & 82.50 & 83.53 & 82.50 & 1.01  & 12.23 \\
EfficientNet-B4 & 72.50 & 75.05 & 72.50 & 1.57  & 19.34 \\
EfficientNet-B5 & 84.00 & 84.74 & 84.00 & 2.44  & 30.39 \\
EfficientNet-B6 & 84.00 & 84.26 & 84.00 & 3.47  & 43.04 \\
\textbf{EfficientNet-B7} & \textbf{86.00} & \textbf{86.32} & \textbf{86.00} & \textbf{5.32}  & \textbf{66.35} \\
\hline
\end{tabular}
\end{table}

\begin{figure}[H]
  \begin{center}
    \includegraphics[width=14cm]{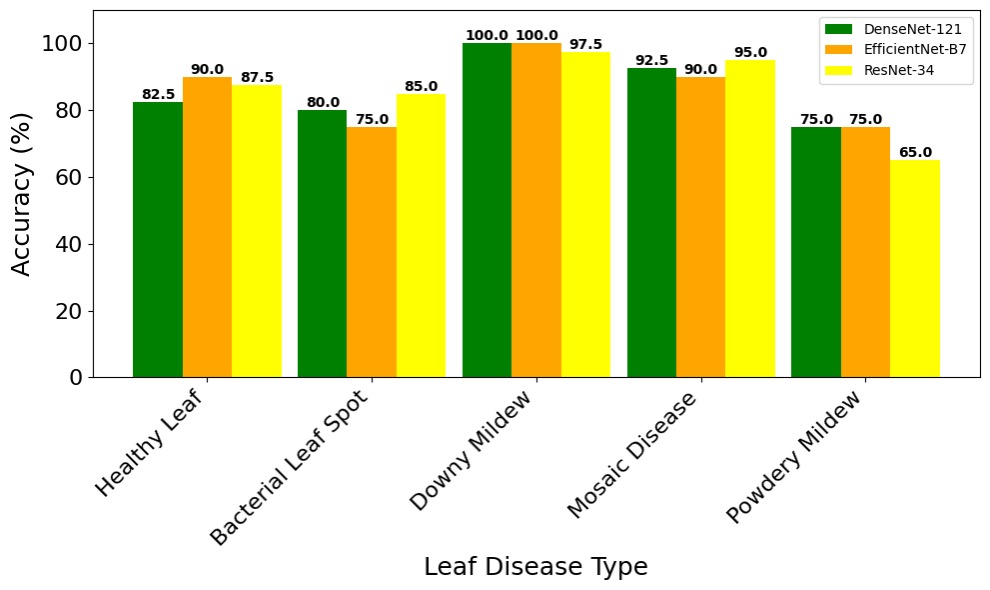} 
  \end{center}
  \caption{Evaluation of the top three models on the test data for different types of pumpkin leaf diseases.}
  \label{fig7}
\end{figure}

Fig. \ref{fig7} presents the test performance of the top three models across different pumpkin leaf diseases. For bacterial leaf spot and mosaic disease, ResNet-34 achieved the highest accuracy at 90\% and 95\%, respectively. EfficientNet-B7 excelled in the healthy leaf class with 90\% accuracy. For downy mildew and powdery mildew, EfficientNet-B7 and DenseNet-121 performed equally well, achieving 100\% and 75\% accuracy, respectively. Considering both accuracy and computational complexity, DenseNet-121 emerges as the most optimal model. Among the disease classes, powdery mildew proved the most difficult to identify, while downy mildew was the easiest.

\begin{figure}[H]
  \begin{center}
    \includegraphics[width=11cm]{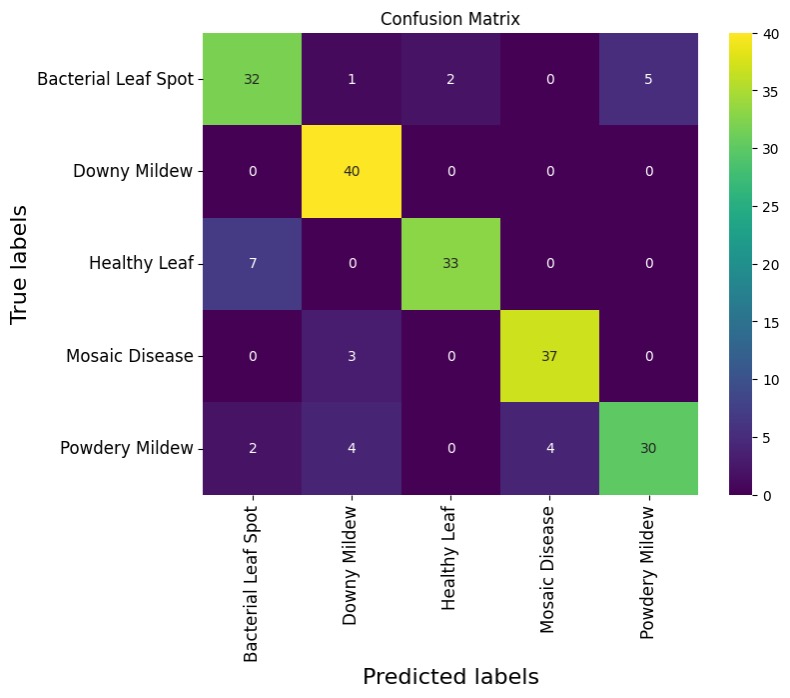} 
  \end{center}
  \caption{Confusion matrix displaying true and predicted pumpkin leaf disease class distributions on test data, generated by DenseNet-121.}
  \label{fig8}
\end{figure}

Fig. \ref{fig8} presents the confusion matrix for DenseNet-121, highlighting areas of misclassification. For example, bacterial leaf spot was often confused with powdery mildew, as both present similar leaf spots. Healthy leaf images were occasionally misclassified as bacterial leaf spot. There were also instances of overlap between downy mildew, powdery mildew, and mosaic disease, potentially due to images exhibiting symptoms of multiple diseases.

\section{Conclusion}\label{S4}
In this study, we conducted a comprehensive analysis of state-of-the-art CNN models for classifying pumpkin leaf diseases. Using a publicly available dataset, we explored the performance of several CNN architectures, including ResNet, DenseNet, and EfficientNet. Our experimental results demonstrated that different models excel in different disease classes, with ResNet-34 performing best for bacterial leaf spot and mosaic disease, and EfficientNet-B7 showing superior performance for the healthy leaf class. While, DenseNet-121 emerged as the most optimal model when balancing both accuracy and computational complexity.

Despite achieving high accuracy in certain classes, challenges remain, particularly with the recognition of powdery mildew, which proved more difficult to classify. This highlights the need for further refinement of models or the incorporation of additional features to improve overall performance. The findings of this study can help guide future research in the automated identification of plant diseases and contribute to enhancing agricultural productivity through better disease management practices.
Future work could focus on testing additional architectures, developing ensemble-based systems that combine predictions from various CNN models, or exploring other plant species to expand the applicability of the developed models.

\section*{Declaration of Competing Interest}
The authors declare that they have no known competing financial interests or personal relationships that could have appeared to influence the work reported in this paper.


\bibliography{sample}

\end{document}